\documentclass[runningheads]{cl2emult}

\usepackage{makeidx}  
\usepackage{graphicx} 
\usepackage{subeqnar} 
\usepackage{multicol} 
\usepackage{cropmark} 
\usepackage{eso}      
\makeindex            

\begin{document}
\title*{Exploring the formation and evolution of massive ellipticals 
with Extremely Red Objects}
%
%
%
%
\titlerunning{Extremely Red Objects}
%
\author{Andrea Cimatti}
\authorrunning{Andrea Cimatti}
%
%
\institute{Osservatorio Astrofisico di Arcetri, Largo E. Fermi 5,
I-50125, Firenze, Italy}

\maketitle              

\begin{abstract}
Extremely Red Objects (EROs) provide the important possibility to
shed light on the formation and evolution of the present-day 
massive ellipticals. On one hand, they allow to select $z>1$ 
old passively evolving spheroidals and to compare their abundance 
with the predictions of galaxy formation scenarios. On the other
hand, they provide the possibility to find dust obscured starbursts, 
a fraction of which may trace the formation of proto-ellipticals
at $z>2$. In this paper, the most recent results on EROs are reviewed 
and the main implications discussed.
\end{abstract}

\section{Introduction}

A fraction of the galaxies selected in the near-infrared show very
red colors (e.g. $R-K>5$). The first cases were serendipitously
discovered by \cite{elston}, and even more extreme cases 
were found by \cite{hr94}. Such galaxies are known as Extremely
Red Objects (EROs), and the most recent surveys demonstrated that 
they form a substantial field population \cite{tho,daddi1} ( 
$\sim$0.5 EROs arcmin$^{-2}$ for $R-K>5$ and $K<19$), whereas
EROs are often found in excess around high-$z$ radio-loud 
AGN \cite{c00,mc00}. Because of the ``age -- dust'' 
degeneracy, the red colors are consistent with EROs being $z>1$ 
old passively evolving ellipticals as well as star-forming galaxies 
and AGN strongly reddened by dust extinction. Because of their faintness,
the observation of EROs is very challenging (sometimes unfeasible) even 
with 10m-class telescopes. However, recent spectroscopy demonstrated 
that ellipticals and starbursts are indeed present in the ERO 
population \cite{gd96,spi,c99,soi,liu}. The key question is then to 
derive the relative fractions of
both galaxy types in order to exploit the stringent clues that EROs
can place on the formation and evolution of elliptical galaxies
and on the abundance of dust obscured systems at high-$z$.

\section{EROs as old ellipticals}

The question on the origin of the present-day massive spheroidals 
is one of the most debated issues of structure formation and
galaxy evolution \cite{renz}. 
The fundamental question is to understand if ellipticals formed at 
early cosmological epochs (e.g. $z>2-3$) through a relatively short 
episode of intense star formation followed by a passive evolution to 
nowadays, or if they built up through the merging of pre-existing 
disk galaxies taking place mostly at $z<1$, thus making massive 
spheroidals at $z>1$ rare objects \cite{k96,bau}. 

A direct way to test such galaxy formation scenarios is then to search for 
massive field ellipticals at $z>1$ and to compare their number with the 
above model predictions. Since near-IR light is a good tracer of the galaxy 
stellar mass \cite{gava} (e.g. $10^{11}$ M$_{\odot}$ correspond to 
$18<K<20$ for $1<z<2$ \cite{k98}), ERO surveys are capable to select  
massive passively evolving elliptical candidates. For instance, a color of 
$R-K>5.3$ is expected for $z>1$, $z_{f}>2$, $H_0$=50 km s$^{-1}$ 
Mpc$^{-1}$, $\Omega_0=0.1-1.0$, $Z=Z_{\odot}$, Salpeter IMF (from Bruzual 
\& Charlot 1997 models). Searches based on this approach gave very discrepant 
results and claimed either a strong deficit or a constant comoving density 
of $z>1$ ellipticals \cite{zepf,fra,bar,tot,ben,bro,schade,scod,treu}. 
Since such results were based on small field surveys (typically 1-60 
arcmin$^2$), wide field searches for EROs have been recently made in 
order to overcome the possible effects of the field-to-field variations.

In one of the widest field survey to date \cite{daddi1}, 700 arcmin$^2$ were 
observed to $K<19$, covering a sky area about ten times larger than the 
previous surveys and providing a complete sample of about 400 EROs with 
$R-K>5$. The main results of such a survey were the detection of strong
angular clustering of EROs (an order of magnitude larger than that
of field galaxies), and the solid estimate of their surface density. 
Such results were recently confirmed by \cite{mc2}. 
The observed clustering can easily explain the previous discrepant 
results as due to strong ``cosmic variance'' effects, and it suggests 
that most EROs with $K<19$ are ellipticals rather than dust reddened 
starbursts 
(see \cite{daddi1} for more details). If this is the case, the spatial 
correlation length $r_0$ of $z\sim0.9-1.5$ ellipticals is in the range of 
6-10$h^{-1}$ Mpc comoving \cite{daddi3,mc2}. 

Concerning the comparison with galaxy evolution scenarios, 
even in a conservative case where 70\% of EROs are ellipticals, their 
observed surface density is in good agreement with the predictions 
of passive evolution (Fig. 1), suggesting that most field ellipticals 
were fully assembled at least by $z=2.5$ \cite{daddi2}. This result does 
not imply that the formation of massive spheroidals occurred necessarily 
through an old fashioned ``monolithic collapse'' \cite{eggen}, but it 
simply constrains the epoch when the formation took place, and it implies 
that, if ellipticals formed through merging, this occurred mostly at $z>2.5$. 

\begin{figure}
\centering
\includegraphics[width=1.1\textwidth, height=14cm]{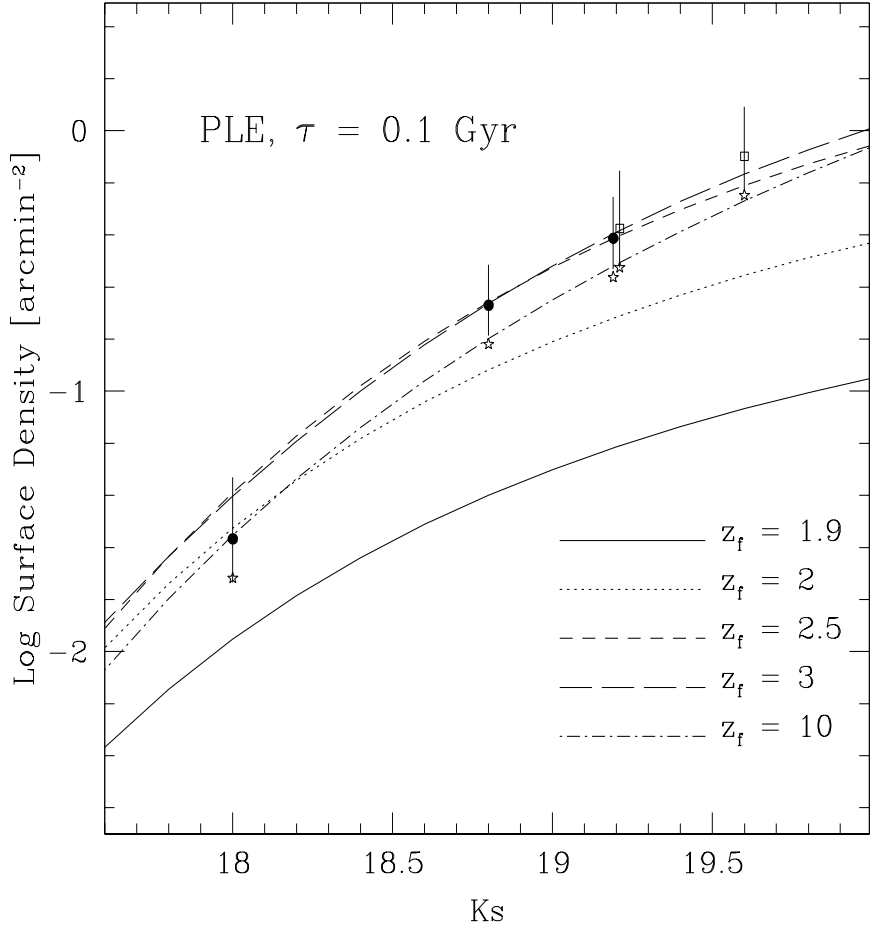}
\noindent
\caption{The filled symbols show the observed surface density of EROs 
with $R-K>5.3$ (corresponding to select passively evolving ellipticals 
at $z>1$, see text) derived from the wide field surveys of \cite{tho,daddi1}.
The error bars take into account the non-poissonian fluctuations due
to the strong angular clustering discovered by \cite{daddi1}.
The curves show the predicted densities in case of pure luminosity 
evolution (PLE) for different formation redshifts adopting the Marzke
et al. (1994) local luminosity function of ellipticals and the Bruzual \&
Charlot (1997) spectral synthesis models with Salpeter IMF and solar 
metallicity (see \cite{daddi2} for more details). The starred symbols 
show the observed density in case 70\% of EROs are passively evolving
elliptical galaxies. The good agreement between the observed and the
predicted surface densities suggests that PLE models cannot be discarded
and that most ellipticals formed at $z>2.5$.
}
\label{eps1}
\end{figure}

Follow-up observations are needed to verify that most EROs are
ellipticals. Despite the observational difficulties due to their 
faintness, the 10-m class telescopes are spectroscopically 
confirming that a substantial fraction of EROs are $z>1$ passively 
evolving ellipticals with old ages ($\sim 1-4$ Gyr) consistent with 
being formed at remote cosmological epochs \cite{spi,liu,soi,c99,cohen,stern,cim_k20}. The preliminary results of the {\it K20} 
survey \cite{cim_k20} show that emission lines were detected 
in only $\sim$20\% of the EROs with $K<20$ and $R-K>5$ observed 
so far, whereas the remaining fraction is made by spectroscopically
identified ellipticals at $z>1$ and by unidentified objects with no
emission lines which could be either ellipticals at higher $z$ or
dusty star-forming galaxies at $z>1.4$ (i.e. with emission lines
falling out of the observed optical spectral range). 
In addition to spectroscopy, recent HST imaging showed that most 
EROs have morphologies and surface brightness profiles characteristic 
of dynamically relaxed spheroidals, whereas only ~15-30\% have irregular 
or disk-like morphologies \cite{morio,sti} (see also \cite{corbin}).

\section{EROs as dusty starbursts and proto-ellipticals}

The possibility that EROs were dust reddened starbursts with 
high far-IR luminosity was first explored by \cite{c98} who 
detected submm and mm continuum emission from HR10 ($z=1.44$; 
\cite{hr94,gd96}) and showed that such a galaxy is an ultraluminous
infrared galaxy (ULIG; $L_{FIR}>10^{12}$ L$_{\odot}$). Strong CO 
emission was then detected by \cite{and} implying a hydrogen
molecular gas mass of $\sim 10^{11}$ M$_{\odot}$. Such results 
demonstrated that a fraction of EROs are dust obscured ULIGs with 
star formation rates (SFRs) $>>100$ M$_{\odot}$yr$^{-1}$ (see 
also \cite{d99}). 

Besides HR10, other EROs have been detected in the submm, and
all of them displayed properties consistent with being ULIGs 
at even higher redshifts
($z_{estimated}\sim2-5$) and with extreme SFRs up to $\sim$1000 
M$_{\odot}$yr$^{-1}$ \cite{smail99,gear,ivison,dunlop}. 
Preliminary SCUBA results suggest that EROs with significant submm 
emission may be segregated among the objects with the most extreme 
colors \cite{cima_scuba} (e.g. $I-K>6$, $R-K>7$). Although 
preliminary, this seems in broad agreement with the HST results 
showing that EROs with irregular/merging-like morphologies 
(i.e. consistent with being dusty starbursts) have the tendency 
to have the reddest colors \cite{morio}. If confirmed, this 
would allow to disentangle dusty EROs from ellipticals on the 
basis of their colors.

If a fraction of massive ellipticals formed at early cosmological 
epochs through a short-lived starburst phenomenon, such a phase 
may have occured in a dusty environment where the UV light from 
OB stars is absorbed by the dust grains and re-emitted in the 
rest-frame far-infrared, making them weak in the observed-frame 
optical and bright in the submillimeter (\cite{granato} and
references therein). It is then intriguing to speculate whether the
dusty EROs which have properties consistent with being ULIGs 
at $z>2$ represent proto-ellipticals seen during their
formation episode. A circumstantial evidence in favour of this 
comes from the similarity between the estimated
comoving density of submm-selected galaxies with $S_{850\mu m}>10$ 
mJy (which often have ERO counterparts) and that of present-day
ellipticals with $L\geq4L^{\ast}$ \cite{dunlop}.

\section{Conclusions and future work}

The most recent results suggest that $z\sim1-1.5$ ellipticals are probably 
the dominant population among the EROs selected at $K\sim$18-20, whereas
a minor, although significant, fraction is made by dust enshrouded
ULIGs over a wide range of redshifts ($1<z<5$). Multiwavelength 
observations of complete samples are needed to establish the 
relative fractions of the galaxies forming the ERO population in 
order to constrain the evolution of massive galaxies and of their 
clustering, and to investigate the link between EROs and 
obscured AGN suggested by recent X-ray observations \cite{hasinger}.

\clearpage
\addcontentsline{toc}{section}{Index}
\flushbottom
\printindex


\begin{thebibliography}{99}
%
\addcontentsline{toc}{section}{References}

\bibitem{elston}Elston R., Rieke G.H., Rieke M.J. 1988, ApJ, 331, L77
\bibitem{hr94}Hu, E. M.; Ridgway, S. E. 1994, AJ, 107, 1303
\bibitem{tho}Thompson D. et al. 1999, ApJ, 523, 100 
\bibitem{daddi1}Daddi E. et al. 2000, A\&A, 361, 535 
\bibitem{c00}Cimatti A. et al. 2000, MNRAS, 318, 453
\bibitem{mc00}Chapman, S. C.; McCarthy, P. J.; Persson, S. E. 2000, AJ,
120, 1612
\bibitem{gd96}Graham J.R., Dey A. 1996, ApJ, 471, 720
\bibitem{spi}Spinrad H. et al. 1997, ApJ, 484, 581
\bibitem{c99}Cimatti A. et al.  1999, A\&A, 352, L45 
\bibitem{soi}Soifer B.T. 1999, AJ, 118, 2065 
\bibitem{liu}Liu M.C. et al. 2000, AJ, 119, 2556
\bibitem{renz}Renzini A., Cimatti A. 2000, in ``The Hy-Redshift Universe:
Galaxy Formation and Evolution at High Redshift", ed. A.J. Bunker \&
W.J.M. van Breugel, A.S.P. Conf. Series Vol. 193, in press (astro-ph/9910162) 
\bibitem{k96}Kauffmann G. 1996, MNRAS, 281, 487 
\bibitem{bau}Baugh C.M., Cole S., Frenk C.S. 1996, MNRAS, 283, 1361
\bibitem{gava}Gavazzi G., Pierini D., Boselli A. 1996, A\&A, 312, 397
\bibitem{k98}Kauffmann G., Charlot S. 1998, MNRAS, 297, L23 
\bibitem{zepf}Zepf S.E. 1997, Nature, 390, 377
\bibitem{fra}Franceschini A. et al. 1998, ApJ, 506, 600
\bibitem{bar}Barger A.J. et al. 1999, AJ, 117, 102
\bibitem{tot}Totani T., Yoshii J. 1997, ApJ, 501, L177
\bibitem{ben}Benitez N. et al. 1999, ApJ, 515, L65
\bibitem{bro}Broadhurst T.J., Bouwens R.J. 1999, ApJ, 530, L53 
\bibitem{schade}Schade D. et al. 1999, ApJ, 525, 31 
\bibitem{scod}Scodeggio M., Silva D.R. 2000, A\&A, 359, 953
\bibitem{treu}Treu T., Stiavelli M. 1999, ApJ, 524, L27
\bibitem{daddi3}Daddi E. et al., these proceedings
\bibitem{mc2}McCarthy P.J. et al., these proceedings
\bibitem{daddi2}Daddi E., Cimatti A., Renzini A. 2000, A\&A, 362, L45
\bibitem{eggen}Eggen, O. J.; Lynden-Bell, D.; Sandage, A. R. 1962, ApJ, 
136, 748
\bibitem{cohen}Cohen J.G. et al. 1999, ApJ, 512, 30
\bibitem{stern}Stern D. et al., these proceedings
\bibitem{cim_k20}Cimatti et al., in preparation (see http://www.
arcetri.astro.it/$\sim$k20/)
\bibitem{morio}Moriondo G., Cimatti A., Daddi E. 2000, A\&A, in press 
\bibitem{sti} Stiavelli M., Treu T. 2000, astro-ph/0010100 
\bibitem{corbin}Corbin M.R. et al. 2000, AJ, 120, 1209
\bibitem{c98}Cimatti A., Andreani P., R\"ottgering H., Tilanus R. 1998, 
Nature, 392, 895
\bibitem{and}Andreani P., Cimatti A., Loinard L., R\"ottgering H.J.A. 2000,
A\&A, 354, L1
\bibitem{d99}Dey A. et al. 1999, ApJ, 519, 610
\bibitem{smail99}Smail I. et al. 1999, MNRAS 308, 1061
\bibitem{gear}Gear W.K. et al. 2000, MNRAS, in press (astroph/0007054)
\bibitem{ivison}Ivison R.J. et al. 2000, ApJ, 542, 271
\bibitem{dunlop}Dunlop J.S. 2000, astroph/0011077
\bibitem{cima_scuba}Cimatti et al., in preparation
\bibitem{granato}Granato G.L. 2000, astro-ph/0006408
\bibitem{hasinger}Hasinger et al., these proceedings
\end{thebibliography}
\end{document}